\documentclass[amsfonts,showpacs,tightenlines,aps,12pt,floatfix]{revtex4}
\usepackage{graphics}
\begin{document}

\title{Many-body approach to proton emission and the role of spectroscopic
factors}
\author{Jim Al-Khalili$^1$}\email{J.Al-Khalili@surrey.ac.uk}
\author{Carlo Barbieri$^2$}\email{barbieri@triumf.ca}
\author{Jutta Escher$^{2,3}$}\email{escher1@llnl.gov}
\altaffiliation[Permanent address: ]
{Nuclear Theory and Modeling Group, Lawrence Livermore National Laboratory,
P.O.\ Box 808, L-414 Livermore, CA 94551, U.S.A.}
\author{Byron K.\ Jennings$^2$}\email{jennings@triumf.ca}
\author{Jean-Marc Sparenberg$^2$}\email{jmspar@triumf.ca}
\affiliation{$^{1}$Department of Physics, University of Surrey, Guildford GU2 7XH, UK,}
\affiliation{$^2$TRIUMF, 4004 Wesbrook Mall, Vancouver, BC, Canada V6T 2A3,}
\affiliation{$^3$Lawrence Livermore National Laboratory,
P.O.\ Box 808, L-414, Livermore, CA 94551, U.S.A.}
\date{\today}
\begin{abstract}
The process of proton emission from nuclei is studied by utilizing the
two-potential approach of Gurvitz and Kalbermann in the context of the
full many-body problem.  A time-dependent approach is used for
calculating the decay width.  Starting from an initial many-body
quasi-stationary state, we employ the Feshbach projection operator 
approach and reduce the formalism to an effective one-body 
problem.
We show that the decay width can be expressed in terms of a one-body
matrix element multiplied by a normalization factor.
We demonstrate that the traditional interpretation of this
normalization as the square root of a spectroscopic factor
is only valid for one particular choice of projection operator.
This causes no problem for the calculation of the decay width in a 
consistent microscopic approach, but it leads to ambiguities in the 
interpretation of experimental results.
In particular, spectroscopic factors extracted from a comparison of the 
measured decay width with a calculated single-particle width may be
affected. 
\end{abstract}


\maketitle

\section{Introduction}
One of the classic problems in quantum mechanics is that of tunneling
through a classically forbidden region or, more specifically, the decay
of a quasi-stationary state to the continuum. In nuclear physics, this
manifests itself in the processes of $\alpha$-decay in heavy nuclei and
proton emission by proton drip-line nuclei. Of particular current interest
are the lifetimes of proton emitters, especially in the lighter
region of the nuclear chart, and the implications of this in nuclear
astrophysics.

Over the years, a number of different theoretical approaches have been
used to describe the decay process in nuclear physics, either by means
of perturbation theory of decaying states or by time reverse study of
resonance states via scattering theory~%
\cite{jackson:77,gurvitz:03,mang:64,thomas:54,arima:74,vogt:96}.
Some authors solve the
time-dependent problem while others use a stationary picture and make
use of approximation methods such as the distorted-wave Born approximation
or the semi-classical Wentzel-Kramers-Brillouin
approach to evaluate the width~\cite{aberg:97}.
 Other more accurate methods, such as
R-matrix theory, are sometimes very sensitive to the channel radius giving
dramatic variation in the calculated widths
\cite{mang:64,thomas:54,arima:74,vogt:96}.
The method of  Gurvitz and Kalbermann~\cite{gurvitz:87,gurvitz:88,gurvitz:03},
also known as the two-potential approach (TPA)~\cite{jackson:77,aberg:97},
is based on splitting the barrier potential
into an interior and an exterior components. The inner potential binds
the particle, which can then be described by a bound eigenstate of the
relative Hamiltonian, while the outer potential acts as a perturbation
that converts it into a quasi-stationary state (a wave packet), which can
decay. 

An important shortcoming of all the above approaches, however, and in
common with the descriptions of so many nuclear processes, is 
the
approximate treatment of the many-body structure effects.  In most
descriptions of the proton-emission process the initial 
($A+1$)-body
wave function is written as a product of an $A$-body 
wave function, 
describing the daughter nucleus, and the proton's 
single-particle
wave function.  The decay width is then written in the 
form of a single-particle
width multiplied by a spectroscopic factor, 
which contains the many-body
information of the system.  This 
procedure, however, makes various
assumptions about the relationship 
between the many-body problem and
the effective one-body problem that 
have to be tested.
 In this work we consider the TPA of  Gurvitz and Kalbermann  and extend it
to properly account for the many-body correlations.

The standard reduction from a many-body problem 
to an effective
one-body picture has been revisited in a recent study 
of radiative proton
capture~\cite{escher:01}.  The work focused on 
one-body overlap
functions and their associated equations of motion. 
The
one-body overlap functions are obtained by integrating the 
product of the
wave functions for an ($A+1$)-body system and its 
$A$-body subsystem
over the coordinates of the latter.
While the 
overlap functions are unambiguously defined, it was demonstrated
in 
Ref.~\cite{escher:02} that useful `auxiliary' one-body functions can 
be
defined in several different ways.  Naturally, the associated 
equations of
motion differ for the three approaches considered in 
Ref.~\cite{escher:02}.
In the current work, we derive expressions for 
the proton decay width using
two of the three approaches mentioned.
The resulting decay widths have formally the 
same structure in both approaches, but the
overall normalization 
factors differ.  Only one of these normalization factors
can be 
interpreted as the square root of a spectroscopic factor.  This 
has
consequences for the interpretation of experimental results and 
in particular
for the determination of spectroscopic factors from 
decay widths.

We start, in Sec.\ \ref{sec:II}, with the time-dependent Schr\"{o}dinger
equation and follow the standard theory of decaying
states~\cite{goldberger:64}. This method is briefly compared with the
scattering approach to decay problems.  In Secs.\ \ref{sec:proj} and
\ref{sec:pert} we use projection operator techniques and perturbation theory to
derive an expression for the decay width in terms of the imaginary part of the
pole in the Green's function matrix element. In Sec.\ \ref{sec:H0}, we describe
the two alternative routes for reducing the many-body problem to an effective
one-body problem.  We compare the resulting two expressions for the width in
Sec.\ \ref{sec:two-expr}.  An alternative expression for the decay width is
given in Sec.\ \ref{sec:alt-expr} which shows more clearly the relation of the
width to the spectroscopic factor.

\section{Proton emission formalism}
\label{sec:II}

In this section we begin to recast the formalism of Gurvitz and Kalbermann 
in a form which is more  convenient for our purposes.
The approach of Refs.\ \cite{gurvitz:87,gurvitz:88} starts with a 
square-integrable wave
function, $|\psi_0\rangle$, which corresponds to the quasi-bound nucleus whose
decay we are interested in.  The initial wave function is close to the
resonance state in the nuclear interior but decays rather than
oscillates in
the exterior region. This wave function cannot be an eigenstate of the
full Hamiltonian or it would have a trivial time dependence and no decay
would take place. Taking $|\psi_0\rangle$ as the wave function at $t=0$, we
follow its time evolution using the time-dependent Schr\"odinger equation:
\begin{eqnarray}
i \hbar \frac{\partial}{\partial t} |\psi(t)\rangle = H |
                   \psi(t) \rangle.
\label{one}
\end{eqnarray}
This initial value problem is solved by using the one-sided Fourier transform
(sometimes called a Laplace transform).  We obtain
\begin{eqnarray}
   -i\hbar |\psi_0\rangle + E |\tilde \psi(E)\rangle &=& H
                   |\tilde\psi(E)\rangle,
\end{eqnarray}
where $|\tilde \psi(E)\rangle = \int_0^\infty dt\;e^{i(E+i\epsilon)t/\hbar} |
\psi(t) \rangle$ and $\epsilon$ is a positive infinitesimal real number.
Solving for $|\tilde \psi(E)\rangle$ we have
\begin{eqnarray}
|\tilde \psi(E)\rangle&=&\frac{i\hbar}{E-H+i\epsilon}|\psi_0\rangle.
\end{eqnarray}
The probability amplitude for the nucleus remaining in the initial state after
a time $t$ is given by
\begin{eqnarray}
\langle\psi_0|\psi(t)\rangle &=& \frac{i}{2\pi}\int_{-\infty}^\infty dE\;
        e^{-iEt/\hbar} \langle\psi_0|\frac{1}{E-H+i\epsilon} |
\psi_0\rangle,
\label{decay}
\end{eqnarray}
which is obtained by taking the matrix element of the previous equation with $
\langle \psi_0 | $ and carrying out the inverse Fourier transform. This is the
Fourier transform of one particular matrix element of the many-body Green's
function.  In general the Lehmann representation of the latter contains
contributions from many poles.  However, the overlap $\langle \psi_0 | \psi (t)
\rangle$ takes a simple form if the right-hand side of Eq.~(\ref{decay}) is
dominated by the contribution of only one pole. In this case the decay rate can
be extracted from the imaginary part of the pole location and we obtain a
simple exponential function which describes the decay of the initial state.
Thus the initial state $|\psi_0\rangle$ should be chosen to minimize the
contributions from other poles.

The same result can be found by considering the related scattering problem.
A quasi-bound state can also be thought of as a resonance in the scattering
amplitude.  We consider scattering of a proton off the $(A-1)$-body system and
we define a t-matrix, $T$, through the equation
\begin{eqnarray}
T a^\dag(k) |\psi_{A-1}\rangle = V |\psi_{A}\rangle,
\end{eqnarray}
where $a^\dag(k)$ is the creation operator for a particle with momentum
$k=\sqrt{2 m E/\hbar^2}$ and $m$ is the reduced mass of the system.
By standard techniques it can be shown that
\begin{eqnarray}
T &=& V + VG_0T\label{tmatrix}\\
   &=& V + VGV,
\end{eqnarray}
where $G_0 = 1/(E-H_0+i \epsilon) $, $G=1/(E-H+i \epsilon)$, and
$H_0$ describes the free motion of the ejected particle with respect
to the
final state of the $(A-1)$-body system, i.e., $H_0 a^\dag(k)
|\psi_{A-1}\rangle = E a^\dag(k) |\psi_{A-1}\rangle$ and  $H=H_0+V$.
We observe that the poles of the t-matrix are given by the poles of 
$V$ and $G$.
Assuming $V$
has no nearby pole we see that the poles of $T$ are just those of the many-body
Green's function. The width of the state is then given by the imaginary part of
the pole location as in the previous case.

In principle, the potential $V$ is the sum of the interactions of the $A$-th
particle with each of the particles in the $(A-1)$-body system.  In nuclear
physics this is generally approximated by a nucleon-nucleus optical
potential. This approach, however, has the disadvantage of loosing track of the
Pauli exchange correlations and other many-body effects. For bound states one
of the principal effects is included through the use of the spectroscopic
factor. However for scattering states there is, strictly speaking, no
spectroscopic factor. Using Eq.~(\ref{decay}) as the starting point we can
formally take the many-body effects into account while deriving an effective
one-body equation.

\section{Projection operator formalism}
\label{sec:proj}

The main ingredient of Eq.~(\ref{decay}) is the matrix 
element of the Green's function,
$M=\langle\psi_0|\frac{1}{E-H+i\epsilon} | \psi_0\rangle$. 
 Its expression  can be simplified as done in
Ref.~\cite{escher:02} by using the projection operator formalism.
 We define a projection operator $P=| \psi_0\rangle\langle\psi_0|$
and the complementary operator $Q=1-P$. The equation for the Green's
function can be written as
\begin{eqnarray}
(E - H)\frac{1}{E-H+i\epsilon}= (E-H)G = 1.
\end{eqnarray}
Acting on the left by $P$ or $Q$ and on the right by $P$ we get the two
equations
\begin{eqnarray}
P(E-H)(P+Q)GP&=&P,\nonumber\\
Q(E-H)(P+Q)GP&=&0.
\end{eqnarray}
Solving the second equation for $QGP$ and substituting this into the 
first equation
we have
\begin{eqnarray}
  \left(E- PHP - PHQ\frac{1}{E-QHQ+i\epsilon}QHP \right) PGP = P.
\end{eqnarray}
Taking the matrix element of this equation with $| \psi_0\rangle$ and using the
explicit form for $P$ we find
\begin{eqnarray}
\left [E - \langle\psi_0| \left( H + HQ\frac{1}{E-QHQ+i\epsilon}QH \right)  |
\psi_0\rangle \right]
\langle\psi_0| G | \psi_0\rangle = 1.
\label{eq:green}
\end{eqnarray}
Since the left-hand side of this equation is a product of two terms, 
the poles of
$\langle\psi_0| G | \psi_0\rangle$ must coincide with 
the zeros of the multiplying
factor.  Thus the poles of the Green's 
function are given by
\begin{eqnarray}
E=\langle\psi_0| \left( H + HQ\frac{1}{E-QHQ+i\epsilon}QH \right)  |
\psi_0\rangle \; ,
\label{eq:zero}
\end{eqnarray}
or, if we use a spectral representation of $\frac{1}{E-QHQ+i\epsilon}$, by
\begin{eqnarray}
E&=& \langle \psi_0| H | \psi_0\rangle + \int_{-\infty}^\infty dE'\;
\frac{|\langle \psi_0|  H Q|\zeta_{E'}\rangle|^2}{E-E'+i\epsilon},
\label{eq:zero2}
\end{eqnarray}
where $|\zeta_{E'}\rangle$ is the solution of the equation
\begin{eqnarray}
E|\zeta_E\rangle = QHQ |\zeta_E\rangle\label{eq:zeta},
\end{eqnarray}
normalized according to $ \langle\zeta_E' |\zeta_E\rangle 
=\delta(E'-E)$.
Note that $|\psi_0\rangle$ is also an eigenstate of 
$QHQ$, but with
energy $E=0$. This state is excluded from the sum 
(integral) in
Eq.~(\ref{eq:zero2}) by the projection operator.  In fact the only 
role of $Q$ in
this equation is to exclude the discrete state $|\psi_0\rangle$. The
residue, $R$, of the pole of $\langle\psi_0| G | \psi_0\rangle$ is given by
\begin{eqnarray}
R=\left[ 1 - \frac{d}{dE} \langle\psi_0| HQ\frac{1}{E-QHQ+i\epsilon}QH
| \psi_0\rangle \right]^{-1}.
\end{eqnarray}
This follows from a Taylor series expansion of Eq.~(\ref{eq:green}) in $E$. For
the present problem $R$ is close to one since we will choose $|\psi_0\rangle$
to be an
eigenvalue of $H$ inside the nucleus and different from an eigenstate only in a
region where the wave function is exponentially damped.
Where $|\psi_0\rangle$ is an
eigenstate $\langle\psi_0| HQ $ is zero so the contribution from the second
term will be exponentially small.

More insight into Eq.~(\ref{eq:zero}) can be obtained by an alternative
derivation.  Consider the equation $E=\langle \psi_0 |H| \psi_E \rangle /
\langle \psi_0 | \psi_E \rangle$ where $| \psi_E \rangle$ is an eigenstate of
the full Hamiltonian, $H$. The wave function can be written as
$|\psi_E\rangle=(P+Q) |\psi_E\rangle$. Following the standard Feshbach
projection operator technique we can write $Q|\psi_E\rangle =
Q\frac{1}{E-QHQ+i\epsilon}QHP |\psi_E\rangle$.  Using the explicit form of the
projection operator $P=| \psi_0\rangle\langle\psi_0|$ yields
Eq.~(\ref{eq:zero}) immediately.

\section{Perturbative approximation}
\label{sec:pert}

In general, Eq.~(\ref{eq:zero}) is highly non-linear and has many solutions.
Every solution of this equation will give a pole of the Green's function,
however, not every pole of the Green's function will necessarily be found by
using this equation. For example, if $|\psi_0\rangle$ has a definite angular
momentum only poles with that angular momentum can be found with that
particular choice of $ |\psi_0\rangle$.

For the problem at hand, namely proton emission, we are not interested in the
complete complexity of the Green's function.  To obtain the decay width, only
the imaginary part of the pole location of the nearby pole is relevant.  We
define an Hermitian Hamiltonian $H_0$ such that $H_0 |\psi_0\rangle = E_0
|\psi_0\rangle$ with $H=H_0+\delta H$. Expanding the right hand side of
Eq.~(\ref{eq:zero}) about $E_0$ and neglecting terms of order $(E-E_0)^2$ we
obtain
\begin{eqnarray}
  E - E_0 &\approx& \langle\psi_0 | \delta H | \psi_0\rangle + \langle\psi_0|
HQ\frac{1}{E_0-QHQ+i\epsilon}QH | \psi_0\rangle \nonumber\\&&
+ (E-E_0) \left[\frac{d}{dE} \langle\psi_0| HQ\frac{1}{E-QHQ+i\epsilon}QH |
\psi_0\rangle\right]_{E=E_0}\\ &\approx&  R\left(\langle\psi_0 | \delta H |
\psi_0\rangle+ \langle\psi_0| HQ\frac{1}{ E_0 - QHQ+i\epsilon}QH |
\psi_0\rangle\right),
\label{eq:e-e0}
\end{eqnarray}
where the last line has been obtained by solving for $(E - E_0)$ in the first
equation.  This expression for the energy, Eq.~(\ref{eq:e-e0}), contains a
factor of $R$, the residue of pole of $\langle\psi_0| G | \psi_0\rangle$.
Note that $\langle\psi_0| HQ=\langle\psi_0|\delta H Q $ implies that
$R-1$ is of the order of $\delta H^2$.
Since it multiplies a factor of order $\delta H$, it will introduce
terms of order higher than $(E-E_0)^2\approx\delta H^2$
and therefore it can be neglected.
In the following, we thus take $R=1$.

Let us emphasize that $E_0$ is real since it is an the eigenvalue of the Hermitian
operator $H_0$.  The pole location, on the other hand, occurs at a complex
energy. If the width of the state is large, for any reason not just proton
emission, then $E\approx E_0$ does not hold (since $E_0$ is real and the pole
location has a large imaginary part) and the approximation fails. However, when
the width is narrow then it will be possible to chose $| \psi_0 \rangle$ or
equivalently $H_0$ such that perturbation theory is applicable.

The equation for the energy is now linear in $E$ and has the form typical of
perturbation theory.  The explicit connection with perturbation theory can be
made when the unperturbed Hamiltonian is taken to be $H_{00}=H_0+Q\delta HQ$,
and not $H_0$ as one might have expected.  This definition is not only
necessary in order to cast Eq.~(\ref{eq:zero2}) into the form of a perturbation
expansion but at the same time eliminates the problem of
non-compactness~\cite{gurvitz:87,gurvitz:88}. The non-compactness arises since
$H_0$ is chosen in such a manner that it does not go to zero at
infinity. However both $H$ and $H_{00}$ go to zero at infinity.

The states $|\zeta_{E} \rangle$ occurring in the integral in
Eq.~(\ref{eq:zero2}) are eigenstates of $H_{00}$, as is $|\psi_0\rangle$.
In contrast to Eq.~(\ref{eq:zeta}) the state $|\psi_0\rangle$ now occurs
with energy $E_0$ rather than $0$.
We can also rewrite $ \langle\psi_0| H Q |\zeta_{E'}
\rangle$ as $ \langle \psi_0| (H-H_0) Q |\zeta_{E'}\rangle $ or $ \langle
\psi_0| (H-H_{00}) Q |\zeta_{E'}\rangle $ since $ \langle \psi_0|$ is an
eigenstate of both $H_0$ and $H_{00}$ or, equivalently, since $Q$ commutes with
both $H_0$ and $H_{00}$. Using the previously defined $\delta H= H-H_0$,
Eq.~(\ref{eq:zero2}) can be rewritten as
\begin{eqnarray}
E&\approx& E_0 + \langle \psi_0|\delta H | \psi_0\rangle + 
\int_{-\infty}^\infty dE'\; \frac{|\langle
\psi_0| \delta H Q |\zeta_{E'}\rangle|^2}{E_0-E'+i\epsilon}\label{eq:zero3}\\
&\approx& E_0 + \langle \psi_0|\delta H' | \psi_0\rangle + 
\int_{-\infty}^\infty dE'\; \frac{|\langle
\psi_0| \delta H' Q |\zeta_{E'}\rangle|^2}{E_0-E'+i\epsilon},
\label{eq:zero3a}
\end{eqnarray}
where
\begin{eqnarray}
\delta H'= H-H_{00}=P\delta H + \delta HP - P\delta HP.
\label{eq:hp}
\end{eqnarray}
Equation (\ref{eq:zero3}) for $E$ is {\em not} in the form of perturbation
theory since $|\zeta_{E}\rangle$ is an eigenfunction of $H_{00}$ and not
$H_0$. However, Eq.~(\ref{eq:zero3a}) is. The only approximation made in
deriving Eqs.~(\ref{eq:zero3}) and (\ref{eq:zero3a}) was the replacement of $E$
in the denominator on the right hand side by $E_0$, i.e., we assumed that
perturbation theory is valid.  

We emphasize that neither $| \psi_0\rangle$ nor $|\zeta_E\rangle$ is an
eigenstate of the full Hamiltonian $H$ and our derivation depends crucially on
this point. Instead, $| \psi_0\rangle$ is an eigenstate of $H_0$ and $H_{00}$
while $|\zeta_E\rangle$ is an eigenstate of $QHQ$ and $H_{00}$. The fact that
both wave functions are eigenstates of $H_{00}$ allows us to 
interpret the right-hand
side of Eq.~(\ref{eq:zero3a}) as the first few terms of a 
perturbative expansion.

\section{Choice of $H_0$}
\label{sec:H0}

The proper choice of $H_0$ was discussed in a very convincing and direct way by
Gurvitz and Kalbermann \cite{gurvitz:87,gurvitz:88}
for the problem of a single particle in a local potential well.
The reader is referred to those papers for specific details.
The basic idea is to take an $H_0$ such that at infinity
the potential goes to a finite value, larger than $E_0$, rather than to zero.
This causes the state under consideration to be bound,  but if the 
decay width
is small this new
state should be very close to the scattering state. We now generalize this
concept to the case of $A$ interacting particles.

We consider the case where there is one open channel. Asymptotically the wave
function describes a free proton and the bound $(A-1)$-body system. 
Thus we take
\begin{eqnarray}
\delta H = - \int d\mathbf{r} \;a^\dag(\mathbf{r})|\Phi_{A-1}\rangle 
V(\mathbf{r})\langle\Phi_{A-1}|
a(\mathbf{r}),
\end{eqnarray}
where $|\Phi_{A-1}\rangle$ is the ground state of the $(A-1)$-body system,
$\mathbf{r}$ is the relative coordinate of the proton and the 
$(A-1)$-body system.
The integral covers the whole space.
If $V(\mathbf{r})$ is taken to be greater than $E_0$ outside the range
of the nuclear potential, $r_0$,
it will prevent the initial state $| \psi_0 \rangle$ from decaying.
At the same time, we want the perturbing potential $V(\mathbf{r})$ to 
be zero
inside the nucleus so that it does not modify the wave 
function in the interior.
With this ansatz expression (\ref{eq:zero3}) for $E$ becomes
\begin{eqnarray}
E&\approx& E_0 - \int d\mathbf{r}\;
\phi^*_0(\mathbf{r}) V(\mathbf{r}) \phi_0(\mathbf{r}) +
\int_{-\infty}^\infty dE'\; \frac{| \int
d\mathbf{r}\;\phi^*_0(\mathbf{r}) V(\mathbf{r})\phi_{E'}(\mathbf{r}) |^2}
	{E_0-E'+i\epsilon},
\label{eq:zero4}
\end{eqnarray}
where $\phi_0(\mathbf{r}) = \langle\Phi_{A-1}|a(\mathbf{r})| \psi_0\rangle$ is
a spectroscopic amplitude (i.e.\ a one-body overlap function involving bound
many-body states) and $\phi_{E'}(\mathbf{r})=\langle\Phi_{A-1}|a(\mathbf{r})|
\zeta_{E'}\rangle$ is an optical model wave function (cf.\ the discussion in
Refs.~\cite{escher:01,escher:02}). 
The projection operator $Q$ does not have to
be included explicitly since the integral does not include the discrete state
$\phi_0(\mathbf{r})$.

A miracle has occurred here. Due to the choice of $\delta H$, which is
physically motivated, the expression has reduced to an effective one-body
problem. All the many-body aspects of the problem are contained in the one-body
overlap functions, $\phi_0(\mathbf{r})$ and $\phi_{E'}(\mathbf{r})$.  We stress
that the only approximation made so far is that second order perturbation
theory was used to justify replacing $E$ with $E_0$ in
Eq.~(\ref{eq:zero2}). For states with narrow widths this should be
acceptable. We have not assumed that $|\psi_0\rangle$ is a product state or
made any other assumptions regarding its structure.

Equation (\ref{eq:zeta}) for $| \zeta_E\rangle$
can be written as $E|\zeta_E\rangle= H_{00} |\zeta_E\rangle$.
In the neighborhood of $E=E_0$ the Hamiltonian
in this equation for $| \zeta_{E'}\rangle$ can be approximated as
\begin{eqnarray}
\delta H'  &\approx& -\int d\mathbf{r}\;a^\dag(\mathbf{r})|\Phi_{A-1}\rangle
        V_{00}(\mathbf{r})\langle\Phi_{A-1}|a(\mathbf{r}).
\label{eq:hp-ap}
\end{eqnarray}
For an appropriately chosen $V_{00}(\mathbf{r})$, the approximation made here
is the same as that of Eq.~(2.15) in Ref.~\cite{gurvitz:87}.  The basic
argument given there proceeds as follows: the eigenfunction of $H$ at resonance
will be large when the $A$th particle is in the nuclear interior. However the
state $| \zeta_{E'}\rangle$ will be small due to the projection operators in
$QHQ$. It will look like the real state in exterior region but be suppressed in
the interior.  The form of $V_{00}(\mathbf{r})$ will ensure this if we take it
to be $0$ in the exterior region and repulsive (greater than $E_0$) in the
interior region (see Fig.~3 in Ref.~\cite{gurvitz:87}).  This is the opposite
of what we did for $V(\mathbf{r})$ which was large in the exterior and $0$ in
the interior. 

Despite its nice form, Eq.~(\ref{eq:zero4}) is not useful until we specify how
to calculate the functions $\phi_0(\mathbf{r})$ and $\phi_{E'}(\mathbf{r})$.
Following Ref.~\cite{escher:02} we set up one-body equations for these
functions using the Feshbach projection operator technique. We start with the
exact scattering state $|\Phi_{A}\rangle$ and write the equation-of-motion for
the corresponding one-body overlap function (which in this case is the Feshbach
optical model wave function) and see how it is modified when the perturbing
Hamiltonian $\delta H$ is added.  The one-body overlap function for the exact
scattering state is given by \cite{feshbach:58,feshbach:62,block:63,escher:02}
\begin{eqnarray}
E\phi(\mathbf{r}) &=& \int d\mathbf{r}'d\mathbf{r}''\langle\Phi_{A-1}|
  a(\mathbf{r})\left( H + H Q_F \frac{1}{E-
Q_F H Q_F} Q_F H\right) a^\dag(\mathbf{r}'') |\Phi_{A-1}\rangle \nonumber\\ &&
\hspace*{2cm} {\cal N}(\mathbf{r}'',\mathbf{r}')^{-1 } \phi(\mathbf{r}')\\
       &=& \int d\mathbf{r}' {\cal H}(\mathbf{r},\mathbf{r}') \phi(\mathbf{r}'),
\end{eqnarray}
where $P_F = \int d\mathbf{r}d\mathbf{r}'\;a^\dag(\mathbf{r})
|\Phi_{A-1}\rangle{\cal N} (\mathbf{r},\mathbf{r}')^{-1}
\langle\Phi_{A-1}| a(\mathbf{r}')$, $Q_F=1-P_F$ and the norm operator is
${\cal N}(\mathbf{r},\mathbf{r}')=\langle\Phi_{A -1}|
a(\mathbf{r})a^\dag(\mathbf{r}')|\Phi_{A-1}\rangle$.
Since $\delta HQ_F=Q_F\delta H=0$,
it follows that $\phi_0(\mathbf{r})$ satisfies the equation
\begin{eqnarray}
E\phi_0(\mathbf{r})=\int d\mathbf{r}'\left[ {\cal H}(\mathbf{r},\mathbf{r}')
  + {\cal N}(\mathbf{r},\mathbf{r}') V(\mathbf{r}') \right]
\phi_0(\mathbf{r}').
\label{eq:phi0schroed}
\end{eqnarray}
Note the explicit asymmetry of this Hamiltonian. Even in the absence of
inelastic processes it is not Hermitian. An analogous potential in alpha
emission has been strongly criticized \cite{varga:91,lovas:98} in the study of
alpha emission. Two methods of addressing the problem were discussed in
Ref.~\cite{escher:02}. The first is to define a new amplitude
$\bar\phi(\mathbf{r})=\int d\mathbf{r}'\; {\cal N}
(\mathbf{r},\mathbf{r}')^{-1/2} \phi(\mathbf{r}')$ and take
\begin{eqnarray}
\delta H = -\int d\mathbf{r}d\mathbf{r}'d\mathbf{r}''
a^\dag(\mathbf{r})|\Phi_{A-1}\rangle {\cal N}(\mathbf{r},\mathbf{r}')^{-1/2}
\bar V(\mathbf{r}') {\cal N}(\mathbf{r}',\mathbf{r}'')^{-1/2}
\langle\Phi_{A -1}| a(\mathbf{r}'').
\end{eqnarray}
The equation for $\bar\phi(\mathbf{r})$ is then
\begin{eqnarray}
E\bar\phi_0(\mathbf{r}) = \int d\mathbf{r}'\left[
\bar{\cal H}(\mathbf{r},\mathbf{r}') + \delta (\mathbf{r},\mathbf{r}') \bar
V(\mathbf{r}') \right] \bar\phi_0(\mathbf{r}'),
\end{eqnarray}
where $\bar{\cal H}(\mathbf{r},\mathbf{r}')$ is given in
Ref.~\cite{escher:02} as
\begin{eqnarray}
\bar{\cal H}(\mathbf{r},\mathbf{r}') = \int d\mathbf{r}'' d\mathbf{r}'''
{\cal N}(\mathbf{r},\mathbf{r}'')^{-1/2}  { \cal H}(\mathbf{r}'',\mathbf{r}''')
{\cal N}(\mathbf{r}''',\mathbf{r}')^{1/2}.
\label{eq:bh}
\end{eqnarray}
A similar development holds for $\phi_E(\mathbf{r})$
and we obtain the corresponding equation
\begin{eqnarray}
E\bar\phi_E(\mathbf{r}) = \int d\mathbf{r}'\left[
\bar{\cal H}(\mathbf{r},\mathbf{r}') + \delta (\mathbf{r},\mathbf{r}') \bar
V_{00}(\mathbf{r}') \right] \bar\phi_E(\mathbf{r}').
\end{eqnarray}
The equations we find for $\bar\phi_0(\mathbf{r})$ and 
$\bar\phi_E(\mathbf{r})$ are
quite remarkable. As previously noted, 
we have reduced the many-body problem
to an effective one-body 
equation in which all many-body effects are contained in
$\bar{\cal 
H}(\mathbf{r},\mathbf{r}')$.
The perturbing potential is taken to be local in both cases. This is
not really an additional approximation since the justifications given
previously still hold.

Another way of getting a symmetric form for the additional potential
is by using a projection operator given in Ref.~\cite{escher:02}
\begin{eqnarray}
P_F = \int d\mathbf{r} \;[a(\mathbf{r})+a^\dag(\mathbf{r})]
|\Phi_{A-1} \rangle \langle\Phi_{A -1}|
[a(\mathbf{r}) + a^\dag(\mathbf{r})] \; ,
\end{eqnarray}
for which Eq.\ (\ref{eq:phi0schroed}) reduces to the Dyson equation of
many-body theory.
For time reversal invariant states the associated norm operator 
reduces to the unity
operator. Repeating the derivation with this new projection operator we obtain
\begin{eqnarray}
E\phi_0(\mathbf{r}) &=& \int d\mathbf{r}'\left[
{\cal H_M}(\mathbf{r},\mathbf{r}') + \delta (\mathbf{r},\mathbf{r}')
V(\mathbf{r}') \right] \phi_0(\mathbf{r}'), \label{MeanFieldH1}\\
E\phi_E(\mathbf{r}) &=& \int d\mathbf{r}'\left[
{\cal H_M}(\mathbf{r},\mathbf{r}') + \delta (\mathbf{r},\mathbf{r}')
V_{00}(\mathbf{r}') \right] \phi_E(\mathbf{r}'), \label{MeanFieldH2}
\end{eqnarray}
where ${\cal H_M}(\mathbf{r},\mathbf{r}')$ is the mass operator
that occurs in the particle-hole Green's function \cite{escher:02}.
In general there is no simple relation between
${\cal H_M}(\mathbf{r},\mathbf{r}')$ and
$\bar{\cal H}(\mathbf{r},\mathbf{r}')$.
The equations for $\phi(\mathbf{r})$ and $\bar\phi(\mathbf{r})$ are 
now formally
the same.  At this point we would expect the widths 
calculated with these functions to be numerically similar.
 The only 
difference in the two treatments is the form of $V(\mathbf{r})$ and 
$V_{00}(\mathbf{r})$. 
When the argument leading to the justification of perturbation theory
is correct we expect both approaches to work equally well.

The argument for Eq.~(\ref{eq:hp-ap}) can be restated using the one-body
functions defined above.  The resonant state, that is the solution of the exact
Schr\"odinger-like equation $E_0 \phi_0^{\rm true}(\mathbf{r}) = \int
d\mathbf{r}'\; {\cal H}(\mathbf{r},\mathbf{r}') \phi_0^{\rm
true}(\mathbf{r}')$, grows exponentially as we go into the classical forbidden
region under the Coulomb barrier from the outside and the wave function is
exponentially large in the interior. On the other hand, $\phi_E(\mathbf{r})$
coming from $|\zeta_E\rangle$, the solution of the many body-state defined with
the Hamiltonian $QHQ$, is exponentially suppressed due to the orthogonality of
$|\psi_0\rangle$ and $|\zeta_E\rangle$. If the suppression is large (i.e.\ the
state under consideration is narrow) the precise form of $|\zeta_E\rangle$ in
the interior is not important as it is essentially zero. The approximate form
of $H_{00}$ is chosen so the approximate $|\zeta_E\rangle$ is also zero in the
interior.  As we will see later we do not need $|\zeta_E\rangle$ in the
interior but only outside at some radius, $r_0$.

\section{Two expressions for the width}
\label{sec:two-expr}

The decay width is given in terms of the imaginary part of the pole location in
Eq.~(\ref{eq:zero4}).  We find
\begin{eqnarray}
\Gamma_0 &\approx& 2 \pi \left| \int 
d\mathbf{r}\;\phi^*_0(\mathbf{r}) V(\mathbf{r})
\phi_{E_0}(\mathbf{r}) \right|^2
\label{eq:widthc}
\end{eqnarray}
or
\begin{eqnarray}
\Gamma_0 
&\approx&2 \pi \left| \int d\mathbf{r}\;\bar\phi^*_0(\mathbf{r}) \bar
V(\mathbf{r}) \bar \phi_{E_0}(\mathbf{r}) \right|^2,
\end{eqnarray}
depending on whether we use the bar-ed amplitudes or the standard 
overlap 
functions.  As discussed below, we expect the approximations 
leading to the two
equations to be valid simultaneously, so that the resulting two 
expressions for the
width will agree.  We stress again that the 
reduction to an effective one-body problem
is not an approximation 
but emerges rather naturally from the formalism.
In a pure one-body 
problem $\phi_0(\mathbf{r})$ and $\bar{\phi}_0(\mathbf{r})$
would be 
normalized to unity whereas here they are normalized to the factors 
$\sqrt{S_0}$ and $\sqrt{\bar{S}_0}$, respectively.  The 
normalization factors
contain the many-body aspects of the problem. 
For the standard overlap function
$\phi_0(\mathbf{r})$, this 
normalization is the well-known spectroscopic factor
(see also 
Ref.~\cite{escher:02}).
It is useful to define normalized one-body functions 
$\hat\phi_0(\mathbf{r})$ and
$\hat{\bar{\phi}}_0(\mathbf{r})$ via
$\phi_0(\mathbf{r})=\sqrt{S_0} \hat\phi_0(\mathbf{r})$ 
and
$\bar{\phi}_0(\mathbf{r})=\sqrt{\bar{S}_0} 
\hat{\bar{\phi}}_0(\mathbf{r})$
(i.e.\ 
$\int d\mathbf{r}|\hat\phi_0(\mathbf{r})|^2=
\int d\mathbf{r} |\hat{\bar{\phi}}(\mathbf{r})_0|^2=1$).
The wave-functions $\phi_{E_0}(\mathbf{r})$ and 
$\bar\phi_{E_0}(\mathbf{r})$
describe scattering states and are 
normalized asymptotically at large $r$.
If we assume the one-body Hamiltonians ${\cal H}$ and 
${\bar{\cal H}}$ to be local
(more will be said about this later) we 
can take over verbatim the development given 
in
Refs.~\cite{gurvitz:87,gurvitz:88,gurvitz:03}.
The only difference 
between the treatment presented there and the approach shown
here is 
the presence of the factors $S_0$ (for the standard overlap 
functions) and
$\bar S_0$ (for the bar-ed function).
To make the dependence on these normalization factors explicit
we rewrite the last equations as
\begin{eqnarray}
\Gamma_0 &\approx& 2 \pi S_0 \left| \int d\mathbf{r}\;\hat \phi^*_0(\mathbf{r})
V(\mathbf{r})\phi_{E_0}(\mathbf{r})
\right|^2
\label{eq:widths}
\end{eqnarray}
and
\begin{eqnarray}
\Gamma_0 
&\approx& 2 \pi \bar S_0 \left| \int d\mathbf{r}\;
\hat{\bar\phi}^*_0(\mathbf{r})
\bar V(\mathbf{r}) \bar\phi_{E_0}(\mathbf{r}) \right|^2.
\label{eq:widthsbar}
\end{eqnarray}

To further understand the results we peruse one particular line of development.
As previously noted, we take $V(\mathbf{r})$ to be zero inside the nucleus,
$V(\mathbf{r})=0$ for $r \le r_0$.  For $r \ge r_0$, we have $ \int
d\mathbf{r}'{\cal H_M}(\mathbf{r},\mathbf{r}') \phi_E(\mathbf{r}') = E
\phi_E(\mathbf{r})$ since $V_{00}(\mathbf{r})$ vanishes for large $r$. For
definiteness we follow Refs.~\cite{gurvitz:87,gurvitz:88} and take $r_0$ to the
maximum of the potential. For the example of fluorine discussed below,
this corresponds to about 5 fm.
Thus we write the integral in Eq.\ (\ref{eq:widthc}) as
\begin{eqnarray}
I &=&  \int d\mathbf{r}\; \theta(r-r_0)
\phi^*_0(\mathbf{r}) V(\mathbf{r})\phi_{E_0}(\mathbf{r})\\
   &=&  \int d\mathbf{r}d\mathbf{r}'\;\theta(r-r_0)\phi^*_0(\mathbf{r})
\left\{ {\cal H_M}(\mathbf{r},\mathbf{r}') +[- E_0 + V(\mathbf{r}) ]
   \delta(\mathbf{r},\mathbf{r}') \right\}\phi_{E_0}(\mathbf{r}),
 \label{eq:39}
\end{eqnarray}
with a similar equation holding for the bar-ed quantity. Naively one 
might expect this
to vanish and indeed it would if the step function $\theta$ were not there.
In the following we shall consider the case in which the non-locality
in ${\cal H_M}(\mathbf{r},\mathbf{r}')$ is small enough to be properly
described by a local potential and an effective mass~$m_k(\mathbf{r})$.
In this case one has~\cite{ma:83}
\begin{equation}
{\cal H_M}(\mathbf{r},\mathbf{r}')=\left[ 
\nabla \frac{-\hbar^2}{2m_k(\mathbf{r})} \nabla+V_{\cal M}(\mathbf{r})\right]
\delta(\mathbf{r},\mathbf{r}') \; .
\label{eq:MassOp_vs_mk}
\end{equation}
In this equation ${\cal H_M}$ is correctly evaluated at the energy $E_0$
of the scattered state, therefore $m_k(\mathbf{r})$ accounts for the
sole spatial non-locality and  differs from the usual definition of
effective mass~$m^*(\mathbf{r})$~\cite{negele:81,jaminon:89}.
When Eq.\ (\ref{eq:MassOp_vs_mk}) holds, the integral, Eq.~(\ref{eq:39}),  can
be simplified through integration by parts and,
for spherically symmetric potentials $V(r)$, $V_{\cal M}(r)$
and effective mass $m_k(r)$, one obtains
\begin{eqnarray}
I&=& \frac{\hbar^2}{2 m_k(r_0)}
\left[\phi_{\mathrm{r}0}^*(r_0)
\phi_{\mathrm{r}E_0}'(r_0)-
\phi_{\mathrm{r}E_0}(r_0)
\phi'^*_{\mathrm{r}0}(r_0) \right] 
\label{eq:i1} \\
&=& \frac{\hbar^2}{2\bar m_k(r_0)} \left[
\bar\phi_{\mathrm{r}0}^*(r_0)
\bar\phi_{\mathrm{r} E_0}'(r_0)
  - \bar\phi_{\mathrm{r}E_0}(r_0)
\bar\phi'^*_{\mathrm{r}0}(r_0) \right],
\label{eq:i2}
\end{eqnarray}
where the wave functions are assumed to factorize as $\phi(\mathbf
r)=\frac{\phi_{\mathrm r}(r)}{r} Y_\ell^m(\theta,\varphi)$ and $Y_\ell^m(\theta,\varphi)$
are spherical harmonic functions of the angular
variables $\theta,\varphi$.  The prime denotes the derivation with respect to
the radial coordinate $r$.

\begin{figure}
\scalebox{0.5}{\includegraphics{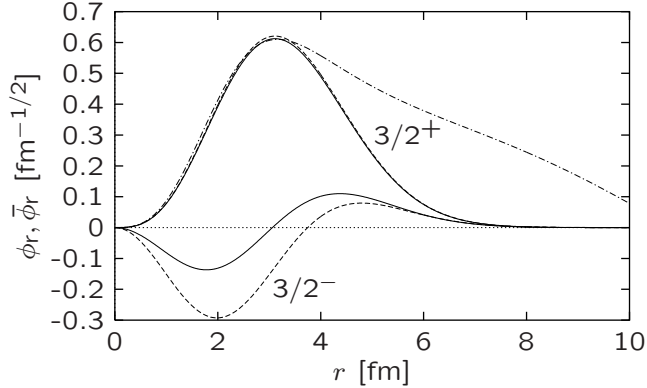}}
\caption{\label{fig:one} Radial part of the one-body overlap functions $\phi$
(solid lines) and of the auxiliary functions $\bar\phi$ (dashed lines)
for the lowest $3/2^-$ and $3/2^+$ states of $^{17}$F,
calculated via the self-consistent Green's function method.
For the $3/2^+$ state, both functions are nearly identical, and a
phenomenological wave function, with a normalization chosen to be close
to that of the microscopic wave functions, is shown for comparison 
(dash-dotted line).}
\end{figure}

In deriving these last two equations we made an additional approximation,
namely that the potentials are local in the vicinity of $r_0$ [or at least the
non-locality is restricted to an effective mass $m_k(r)$]. The two equations
will be simultaneously valid only if the norm operator is local, i.e.\  ${\cal
N}(\mathbf{r},\mathbf{r}')= {\cal
N}(\mathbf{r})\delta(\mathbf{r}-\mathbf{r}')$, in the vicinity of $r_0$. In
this case $\bar m_k(r_0)/m_k(r_0)={\cal N}(r_0)$ and the two equations are
identical.  Since $r_0$ is in the tail of the density distribution, one expects
the norm operator to be unity in its vicinity, which means that the two
expressions for the width should give the same result for realistic models.
This is verified numerically in Fig.~\ref{fig:one}, where the function
$\phi_\mathrm{r}(r)$ (solid lines) and $\bar\phi_\mathrm{r}(r)$ (dashed lines)
are displayed for the $3/2^-$ and $3/2^+$ decaying states of $^{17}$F at
respective excitation energies of 4.64 and 5.00 MeV (the proton threshold is at
0.60 MeV).  The functions are calculated with the self-consistent Green's
function method of Refs.\ \cite{barbieri:01,barbieri:02,barbieri:02t}.
The $3/2^+$ state is well explained by the nuclear mean field approach and
provides a typical example of a strong state, for which $S_0\approx\bar S_0$.
The $3/2^-$ state, on the other hand, is a typical example of a weak state for
which Ref.\ \cite{barbieri:02t} gives $S_0\approx 0.04$ and~$\bar S_0\approx
0.14$.
It is worth to emphasize the difference between these two cases.
For strong states the quenching of the spectroscopic factor is due
both to short-range correlations and to the coupling to other excitations of
the system~\cite{barbieri:02}. Nevertheless they maintain a strong
single-particle structure and the orbital occupancy is of the order of 
unity.
 Weak states, instead, have a more complicated structure and can be seen
as collective excitations on their own rather than having a single-particle
character. As a consequence the one-body spectroscopic factor can be 
one order of magnitude smaller or less. In this  case the functions
$\phi(\mathbf r)$ and $\bar\phi(\mathbf r)$ sample the one-body substructure
in a different way, whence the difference in their normalizations with
$\bar S_0$ larger than~$S_0$~\cite{escher:02}.
For $^{16}$O + p, the nuclear interaction becomes negligible beyond 5 fm
typically and Fig.\ \ref{fig:one} shows that $\phi(\mathbf{r})$
and $\bar\phi(\mathbf{r})$ are equal
beyond this radius, which means that the norm operator is unity.
In principle, there is no problem calculating the width
from a microscopic model:
one may define two different functions $\phi(\mathbf{r})$
and $\bar\phi(\mathbf{r})$,
which have two different normalization factors $\sqrt{S}$ and 
$\sqrt{\bar S}$,
respectively,
but the estimate for the width will be the same with both of them.
In fact, a similar formula would be valid for any amplitude of the form
$\phi^\nu(\mathbf{r})=\int dr'\;{\cal N} 
(\mathbf{r},\mathbf{r}')^{\nu} \phi(\mathbf{r}')$
for arbitrary $\nu$, since ${\cal N}(\mathbf{r},\mathbf{r}')$
is unity around $r_0$.

\begin{figure}
\scalebox{0.5}{\includegraphics{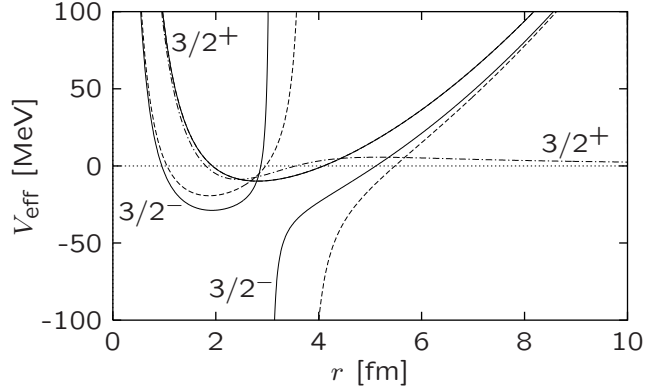}}
\caption{\label{fig:two} Effective (interaction + centrifugal)
potentials corresponding to the wave functions of Fig.\ \ref{fig:one}.
The $3/2^-$ potentials have an $\ell=1$ centrifugal term and a singularity
around 3-4 fm; the $3/2^+$ potentials have an $\ell=2$ centrifugal term and
are regular.  The dash-dotted line indicates the potential 
corresponding to
the phenomenological $3/2^+$ wave function.}
\end{figure}

Let us emphasize that for the actual calculation of the width
by the above formulas,
a microscopic model based on the harmonic oscillator basis, as in
Refs.~\cite{barbieri:02,barbieri:02t} may not be the
best choice since the wave function is not expected to be very precise
in the vicinity of $r_0$.
This is demonstrated in Fig.\ \ref{fig:one},
where the $3/2^+$ resonant wave function of the phenomenological mean field
Woods-Saxon potential of Ref.\ \cite{sparenberg:00b} is shown for comparison.
The agreement with the microscopic wave function is good in the interior
but deteriorates above 4 fm (where the approximation of the potential
by a harmonic oscillator breaks down, see Fig.\ \ref{fig:two}).
The difference between both models is particularly large in this case
since the $3/2^+$ state is wide ($\Gamma_0=1.5$ MeV);
for narrow states, the harmonic oscillator approximation could be
sufficient, a possibility which will be explored elsewhere.
Let us finally stress that the phenomenological mean field potential
of Ref.\ \cite{sparenberg:00b} does not reproduce the $3/2-$ state
because its structure is not well approximated by an $^{16}$O core plus
a proton.

Equations~(\ref{eq:widths}) and (\ref{eq:widthsbar}) 
indicate that there may be
a serious problem in extracting 
spectroscopic factors from measured decay widths.
The standard method 
for determining a spectroscopic factor involves dividing 
an
experimental width by a single-particle width calculated with a 
phenomenological model.
However, it is not clear a priori whether the phenomenological wave functions
are good approximations to $\hat\phi_0(\mathbf{r})$ and $\phi_E(\mathbf{r})$,
or to $\hat{\bar\phi}_0(\mathbf{r})$ and $\bar\phi_E(\mathbf{r})$;
hence, it is not obvious whether dividing the experimental width by 
the result of a single-particle calculation
provides the spectroscopic factor $S_0$ or 
the normalization $\bar S_0$ of the
auxiliary function 
$\bar\phi(\mathbf{r})$ (or the norm of yet another one-body function).
Normally, one assumes in proton emission studies that 
$\hat\phi_0(\mathbf{r})$ and
$\phi_E(\mathbf{r})$ can be equated with the wave functions obtained from
phenomenological potentials  (see for example Ref.~\cite{aberg:97}).
On the other hand, in the context of some alpha emission studies it has been
argued very strongly that $\hat{\bar\phi}_0(\mathbf{r})$ and 
$\bar\phi_E(\mathbf{r})$
correspond to phenomenological wave 
functions~\cite{varga:91,lovas:98} .
If the latter is true then the 
experiments would be sensitive to $\bar S_0$ rather
than to 
$S_0$.
For strong states, with a clear core-plus-particle structure, 
this is mainly a philosophical
issue, since $S_0\approx\bar S_0$ holds.
For weak states, however, $S_0$ and $\bar S_0$ can be 
significantly different
from each other.

We have attempted to resolve the ambiguity outlined above by 
calculating the effective
local potentials corresponding to
$\phi(\mathbf{r})$ and $\bar\phi(\mathbf{r})$
and comparing them with typical phenomenological potentials.
This is done by inversion of the local (radial) Schr\"odinger equation:
\begin{eqnarray}
V_{\rm eff}(r) = E + \frac{\hbar^2}{2m} \frac{\phi_\mathrm{r}''(r)}
{\phi_\mathrm{r}(r)},
\label{eq:inversion}
\end{eqnarray}
where the effective potential $V_\mathrm{eff}(r)$ is the sum of the interaction
potential (nuclear + Coulomb) and the centrifugal term.  The effective
potentials corresponding to the radial wave functions of Fig.\ \ref{fig:one}
are shown in Fig.\ \ref{fig:two}.  For the strong $3/2^+$ state ($\ell=2$
centrifugal term) the three potentials are in reasonable agreement, except
above 4 fm where the potential extracted from the microscopic functions
asymptotically approaches the harmonic oscillator potential that generated
them.  For the weak $3/2^-$ state ($\ell=1$ centrifugal term) the potentials
deduced from $\phi(\mathbf{r})$ and $\bar\phi(\mathbf{r})$ display a
singularity and are very different from traditional phenomenological
potentials.  This singularity occurs because the zeroes of $\phi_\mathrm{r}(r)$
and $\phi_\mathrm{r}''(r)$ occur at different radii.  This is probably not an
artifact of the model but a real effect and arises from the relative sign of
the 0 and $2\hbar\omega$ contributions to the spectroscopic amplitude.  Let us
remark that Eq.\ (\ref{eq:inversion}) assumes a constant effective mass
$m_k(r)=m$.  We have checked that introducing a realistic effective mass does
not lead to a simpler potential for the weak state; more will be said about
this elsewhere.  Let us finally emphasize that, though their energies are close
to one another, the strong and weak states correspond to very different
potentials, which suggests that a strong energy dependence is a necessary
feature for such potentials (see the discussion in the next section).  In
conclusion, this example shows that constructing reliable phenomenological
local potentials for extracting spectroscopic factors from experimental cross
sections is non-trivial.  Moreover, since the characteristics of the potential
depend strongly on the state it is difficult to determine which normalization
($S_0$ or $\bar S_0$) would be extracted from a comparison with the
experimental data.

\section{Alternative expressions for the width}
\label{sec:alt-expr}

Let us now return to the expression for the decay width and establish
a link with known results.
In Refs.~\cite{gurvitz:87,gurvitz:88,gurvitz:03}
the spatial derivatives were evaluated
using a special form for $V(r)$ and $V_{00}(r)$.
In those references these auxiliary potentials were chosen such that
the total potential for $\phi_{\mathrm{r}0}(r)$ was constant outside a
given radius and the one for $\phi_{\mathrm{r}E_0}(r)$ was constant inside.
In Refs.\ \cite{gurvitz:87,gurvitz:88} the separation radius is at the
maximum of the potential barrier, whereas in Ref.\ \cite{gurvitz:03} the advantage
of using a radius of the order of the nuclear range (our $r_0$) is pointed out.
For this last case, the width is then given by
\begin{eqnarray}
\Gamma_0 \approx 2 \pi \left[\frac{\hbar^2 \alpha}{2 m_k(r_0) }\right]^2
|\phi_{\mathrm{r}0}(r_0)\phi_{\mathrm{r}E_0}(r_0)|^2,
\end{eqnarray}
where $\alpha=\sqrt{2 m_k(r_0) [V(r_0)-E_0]/\hbar^2}$.

An alternative approach is to exploit the Wronskian form of
Eqs.~(\ref{eq:i1}) or (\ref{eq:i2}).
If $V(r)$ is zero for radii less than the outer turning point
and $V_{00}(r)$ zero for radii greater than the inner turning point
then there is a region where $\phi_{\mathrm{r}0}(r)$ and
$\phi_{\mathrm{r}E_0}(r)$ satisfy the same differential equation.
If the potential is local the Wronskian is a constant.
Assuming a thick barrier, there will be a region
where $\phi_{\mathrm{r}0}(r)$ is an irregular Coulomb function,
the regular Coulomb function having decayed away,
and where $\phi_{\mathrm{r}E_0}(r)$ is a regular
Coulomb function, the irregular Coulomb function having decayed away.
All that is required is to determine the proportionality constants.
These have a simple expression for a constant effective mass $m_k(r)=m$:
for $\phi_{\mathrm{r} E_0}(r)$ we have
$\phi_{\mathrm{r} E_0}(r)= \sqrt{2m/\hbar^2 k_0 \pi} F(k_0r)$,
with $k_0=\sqrt{2mE_0/\hbar^2}$,
while for $\phi_{\mathrm{r}0}(r)$ the proportionality constant can
be written as $\phi_{\mathrm{r}0}(r_0)/G(k_0 r_0)$.
Here $F(k_0 r)$ and $G(k_0 r)$ are the regular and irregular
Coulomb functions respectively, the Wronskian of which equals $-k_0$.
The resulting width is then, as in Ref.~\cite{gurvitz:03}, 
\begin{eqnarray}
\Gamma_0 \approx \frac{\hbar^2 k_0}{m} |\phi_{\mathrm{r}0}(r_0)/G(k_0r_0)|^2.
\end{eqnarray}
This can be simplified further.
The wave function $\phi_{\mathrm{r}0}(r)$ is normalized
to the appropriate spectroscopic factor, $S_0$. We can also consider the true
scattering wave function at resonance,
$\phi^{\rm true}_{\mathrm{r}0}(r)$.
In the interior it will behave like $\phi_{\mathrm{r}0}(r)$
while in the exterior region it will behave like $G(k_0r)$.
Normalizing it to $G(k_0r)$ in the exterior region we obtain for the width:
\begin{eqnarray}
\Gamma_0 \approx \frac{S_0 \hbar v_0}{\int_0^{r_\mathrm{t}} dr
|\phi^{\rm true}_{\mathrm{r}0}(r)|^2 },
\label{eq:widthf}
\end{eqnarray}
where $v_0=\hbar k_0/m$ is the asymptotic velocity. The exact value of the
upper limit on the integral is not crucial and we take it to be the outer
turning point (see numerical justification below).

Equation (\ref{eq:widthf}) can also be derived in
a more transparent way (see also Refs.~\cite{breit:59,iliadis:97}).
Consider the Schr\"odinger-like equation $[H(E)-E]\phi(\mathbf{r},E)=0$
for the overlap function.
Since the effective one-body equation we have been considering 
can
depend on the energy we keep an explicit energy dependence in 
the
Hamiltonian. Differentiating this equation with 
respect to
$E$ we find
\begin{eqnarray}
[H(E)-E]\frac{\partial \phi(\mathbf{r},E)}{\partial E} =
\left[1- \frac{\partial H(E) }{\partial E}\right] \phi(\mathbf{r},E).
\end{eqnarray}
Next we multiply by $\phi^*(\mathbf{r},E)$ and integrate up to some 
radius $r_l$. If $H(E)$
is not hermitian then $\phi^*(\mathbf{r},E)$ should be replaced by 
the time reversed
state. If the potential is local in the vicinity of $r_l$ we can 
integrate by parts on the left hand side.
Assuming spherical symmetry, this gives us
\begin{eqnarray}
&&-\frac{\hbar^2}{2 m_k(r_l)}   \left[\phi^*_\mathrm{r}(r_l,E)
\frac{\partial \phi_\mathrm{r}'(r_l,E)}{\partial E} -
\phi'^*_\mathrm{r}(r_l,E) \frac{\partial\phi_\mathrm{r}(r_l,E)}{\partial E}
\right]\nonumber\\ &&\hspace{3cm}=
\int_0^{r_l} dr \phi_\mathrm{r}^*(r,E)
\left[1-\frac{\partial H_\mathrm{r}(E)}{\partial E}\right] 
\phi_\mathrm{r}(r,E),
\end{eqnarray}
where the prime denotes a partial derivative with respect to $r$ and
$H_\mathrm{r}(E)$ is the radial Hamiltonian.
If we take $r_l$ to be outside the range of nuclear force,
$\phi_\mathrm{r}(r,E)$ can be written as
\begin{eqnarray}
\phi_\mathrm{r}(r,E)\mathop{\approx}_{r> r_0}
\cos{\delta(E)}F(kr) + \sin{\delta(E)}G(kr).
\end{eqnarray}
At a narrow resonance the phase shift $\delta(E)$ will be rapidly varying so we
expect that the largest part of the energy dependence will come from the phase
shift and not from the Coulomb functions. The energy dependence of the Coulomb
functions will be minimized if the radius is chosen to be near the 
outside turning point.
For example,
for large $r$ the regular Coulomb function will have a $\sin(kr)$ dependence.
Differentiating
with respect to $E$ will give $ r (dk/dE) \cos(kr)$, which diverges
for large $r$.
As $r$ decreases
to the turning point this asymptotic form for the wave function breaks down.
However a
similar argument using exponentials holds inside the turning points.
Thus near
the outside turning point we have
\begin{eqnarray}
\frac{\partial \phi_\mathrm{r}(r,E)}{\partial E} \approx
\frac{d\delta(E)}{d E}
[-\sin{\delta(E)}F(kr) + \cos{\delta(E)}G(kr)].
\end{eqnarray}
We have checked this relation numerically for Woods-Saxon plus
Coulomb potentials and verified that for widths less then 15 KeV the
error does not exceed 3\%.
That the radius should be chosen near the outer turning point was also
confirmed numerically. The Wronskian relation can now be written as
\begin{eqnarray}
  \frac{\hbar^2 k}{2 m_k(r_\mathrm{t})}\frac{d\delta(E)}{d E} \approx
\int_0^{r_\mathrm{t}} dr \phi_\mathrm{r}^*(r,E) \left[1- \frac{\partial
H_\mathrm{r}(E)}{\partial E}\right] \phi_\mathrm{r}(r,E),
\end{eqnarray}
where we have taken $r_l=r_\mathrm{t}$ to be the outside turning point radius.

At the resonant energy we expect the energy variation of the phase shift to be
a maximum so the resonance energy occurs when $\int_0^{r_\mathrm{t}} dr
\phi^*_\mathrm{r}(r,E) \left[1- \frac{\partial H_\mathrm{r}(E)}{\partial
E}\right] \phi_\mathrm{r}(r,E)$ is a maximum.  At the resonance energy
$d\delta(E_0)/d E = 2/\Gamma_0$, and we have
\begin{eqnarray}
\Gamma_0\approx\frac{ \hbar v_0 }{\int_0^{r_\mathrm{t}} dr
\phi^*_\mathrm{r}(r,E_0) \left[1- \frac{\partial
H_\mathrm{r}(E_0)}{\partial E}\right] \phi_\mathrm{r}(r,E_0)} \; ,
\label{eq:widtha}
\end{eqnarray}
where now $v_0=\hbar k_0/m$ is the asymptotic value of the velocity.  For bound
states the spectroscopic factor can be written as $S_0 = \left[\int_0^\infty dr
\phi^*_\mathrm{r}(r) \phi_\mathrm{r}(r)\right]/\left\{\int_0^\infty dr
\phi^*_\mathrm{r}(r) \left[1- \frac{\partial H_\mathrm{r}(E)}{\partial E}
\right]\phi_\mathrm{r}(r)\right\}$ (see Ref.~\cite{wegmann:69}), which does not depend on the specific
normalization of the overlap function~$\phi(\mathbf{r})$.  By extending this
relation to define the spectroscopic factor for resonant states, we recover
Eq.~(\ref{eq:widthf}).

We finally note that the expression (\ref{eq:widtha})
of the width is independent of the choice of $\phi_\mathrm{r}(r,E)$
or $\bar\phi_\mathrm{r}(r,E)$.
Its denominator can be rewritten as
\begin{eqnarray}
D=\int_0^{r_\mathrm{t}} dr \phi^*_\mathrm{r}(r,E) \left\{
\frac{\partial}{\partial E} \left[E- H_\mathrm{r}(E)\right]\right\}
\phi_\mathrm{r}(r,E).
\end{eqnarray}
The bar-ed amplitudes are defined as
$\bar\phi(\mathbf{r},E)=\int d\mathbf{r}\;
{\cal N} (\mathbf{r},\mathbf{r}')^{-1/2} \phi(\mathbf{r},E)$
and a corresponding expression for the Hamiltonian is
$\bar H(E) = {\cal N}^{1/2} \left[ H(E) - E \right]
{\cal N}^{1/2} + E$.
Thus the denominator is invariant under this transformation,
as well as under a general transformation with an arbitrary power of
$\cal N(\mathbf{r},\mathbf{r}')$.
All that matters is that $H_\mathrm{r}(E)$ and $\phi_\mathrm{r}(r,E)$
are consistent with one another.
These wave functions are phase equivalent and any of them can be used.

\section{Discussion}

We have
embedded the elegant Gurvitz-Kalbermann approach of proton
emission~\cite{gurvitz:87,gurvitz:88,gurvitz:03} into a full
many-body picture. We have reduced the formalism to an effective one-body 
problem and demonstrated that the decay width can be expressed in 
terms of a one-body matrix element multiplied by a normalization factor.
At first
sight, this result agrees with the standard procedure for extracting
spectroscopic factors from measurements via dividing an experimental
width by a calculated single-particle width (see, for example,
Ref.~\cite{aberg:97}).
The present work,
however, clearly demonstrates that this procedure for determining
spectroscopic factors is only valid if the phenomenological potential used 
to generate the single-particle width corresponds to the potential in 
${\cal H}_{\cal M}$ [see Eqs.~(\ref{MeanFieldH1}) and (\ref{MeanFieldH2})].
It is not
a priori clear that this is actually the case.  In fact, the authors of 
Refs.~\cite{varga:91,lovas:98}  (and prior to that the authors
of Ref.~\cite{fliessbach:76}) have argued strongly that the
phenomenological potential approximates the potential
in $\bar{\cal H}$ [see Eq.~(\ref{eq:bh})].  While the studies
of Refs.~\cite{varga:91,lovas:98} were carried out for alpha 
decay, the arguments given there can be carried over to a
description 
of the proton emission process.
Furthermore, Eq.~(\ref{eq:widtha}) 
suggests that $\int_0^{r_t} dr \phi^*_\mathrm{r}(r)
\left[1- \frac{\partial H_\mathrm{r}(E)}{\partial E}\right] 
\phi_\mathrm{r}(r)$ is the appropriate
observable that can be 
extracted from proton emission experiments.

Besides the ambiguity regarding whether the standard spectroscopic factor
or an auxiliary  normalization is extracted from the experimental procedure,
it has been demonstrated that constructing a 
reliable phenomenological potential is non-trivial.  The situation is 
quite complicated since the interaction with the nuclear medium strongly
depends on the initial state of the ejected proton.
For states with a clear core-plus-particle 
structure (i.e., with a large spectroscopic factor),
traditional 
phenomenological potentials seem to provide good approximations to 
the nuclear
mean field and spectroscopic factors can be determined 
from proton emission studies. In this case, the spectroscopic factor
extracted from the experiment can be safely compared to the results
of nuclear many-body calculations (note also
that for 
large $S_0$ values, $S_0$ and $\bar S_0$ are approximately 
equal~\cite{escher:02}
and the distinction between the two approaches 
discussed here becomes irrelevant).

For weak states, which have a more complicated many-body structure, standard
phenomenological potentials do not give a proper approximation to the
nuclear medium, as shown by the radial shape of the 3/2$^-$ states in
Fig.~\ref{fig:two}.  Also, as discussed in Ref.~\cite{escher:02}, the
dependence of the spectroscopic factor on the energy derivative of the
effective one-body Hamiltonian implies that the nuclear medium must be strongly
energy dependent. This feature, which is missing in most phenomenological
optical potentials, is also confirmed by the numerical results displayed in
Fig.~\ref{fig:two}. Thus for weak states simple potential models are probably
not valid for either $\phi(\mathbf{r})$ or $\bar\phi(\mathbf{r})$. States that are neither weak
nor strong will also have to be dealt with on a case by case basis. 

Since $\phi(\mathbf{r})$ and $\bar\phi(\mathbf{r})$ are identical for large radii they are phase
equivalent and elastic scattering experiments cannot distinguish between them.
We conclude that additional experimental input, together with an accurate
derivation of the optical potential based on first principles, is required in
order to resolve the question regarding which one-body Hamiltonian is most
appropriately approximated by a phenomenological model.

\begin{acknowledgments}

The authors thank C.~Chandler for valuable discussions and comments on a
preliminary draft of the paper. Financial support from the Natural Sciences and
Engineering Research Council of Canada (NSERC) is appreciated.
This work was performed in part under the auspices of the U. S. Department of
Energy by the University of California, Lawrence Livermore National Laboratory
under contract No. W-7405-Eng-48.
\end{acknowledgments}


\end{document}